\documentclass[a4paper,12pt]{article} 
\usepackage{amsmath, amsthm, amssymb}
\usepackage{url}

\usepackage{physics}
\usepackage{graphicx,lipsum}

\usepackage[colorlinks,citecolor=red,urlcolor=blue,bookmarks=false,hypertexnames=true]{hyperref}
\usepackage{tikz}
\usetikzlibrary{calc}
\usetikzlibrary{shapes}
\usepackage[autostyle]{csquotes}
\makeatletter

\usepackage[colorlinks]{hyperref}
\usepackage[nameinlink,capitalize]{cleveref}

\newtheoremstyle{named}{}{}{\itshape}{}{\bfseries}{.}{.5em}{\thmnote{#3's }#1}
\theoremstyle{named}

\theoremstyle{definition}

\theoremstyle{remark}

\theoremstyle{conclusion}

\theoremstyle{observation}

\usepackage{booktabs}
\usepackage{xspace}
\usepackage[margin=1.0in]{geometry}

\usepackage{titlesec}

\usepackage{mathtools}

\DeclarePairedDelimiterX{\inp}[2]{\langle}{\rangle}{#1, #2}
\titleformat{\chapter}
  {\Large\bfseries} 
  {}                
  {0pt}            
  {\huge}

  \title{A Robust Optimization Model for Cost-Efficient and Fast Electric Vehicle Charging with L2-norm Uncertainty}

\author{
    Trung Duc Tran\thanks{College of Engineering and Computer Science (Vinuniversity), 24duc.tt@vinuni.edu.vn},  
    Ngoc-Doanh Nguyen\footnotemark[1], 
    Hong T.M. Chu\footnotemark[1], \\
    Laurent El Ghaoui\footnotemark[1],
    Luca Ambrosino\thanks{Politecnico di Torino, luca.ambrosino@polito.it}, Giuseppe Calafiore\footnotemark[2]
}
 \date{}

\begin{document}
\maketitle

\begin{abstract}
In this paper, we propose a robust optimization model that addresses both the cost-efficiency and fast charging requirements for electric vehicles (EVs) at charging stations. By combining elements from traditional cost-minimization models and a fast charging objective, we construct an optimization model that balances user costs with rapid power allocation. Additionally, we incorporate L2-norm uncertainty into the charging cost, ensuring that the model remains resilient under cost fluctuations. The proposed model is tested under real-world scenarios and demonstrates its potential for efficient and flexible EV charging solutions.
\end{abstract}

\section{Introduction}

The rapid growth of electric vehicles (EVs) has introduced significant challenges in minimizing total user costs and reducing charging time. As the EV market expands, ensuring efficient charging management becomes critical to maintaining grid stability and minimizing operational costs while meeting user expectations for fast and cost-effective charging solutions. Numerous studies have highlighted the strain that uncontrolled EV charging can place on local distribution networks, potentially leading to grid overload during peak demand hours (like in \cite{clement2010impact}, \cite{coignard2018clean}, and \cite{cross2016my}).

To address these challenges, researchers have proposed various strategies that focus on optimizing charging schedules, integrating renewable energy sources, and utilizing real-time grid conditions to balance demand. For example, algorithms that adjust charging rates based on grid load have been shown to optimize performance and reduce costs (\cite{chen2014electric},  \cite{khaki2018hierarchical}). Moreover, fast charging remains a crucial factor in user satisfaction, and solutions that reduce charging times while minimizing costs are essential for enhancing the user experience (\cite{lee2016shared}, \cite{rivera2017distributed}).

In addition to managing grid demand, integrating renewable energy sources into EV charging systems offers an opportunity to decarbonize transportation. Several studies have explored the potential of combining solar power with EV charging to smooth demand peaks and reduce the carbon footprint of EVs (\cite{ardakanian2014quantifying}, \cite{denholm2013co}). This approach not only helps mitigate the environmental impact of EVs but also enhances the flexibility and resilience of the grid by leveraging intermittent renewable energy resources (\cite{schuller2015quantifying}, \cite{wu2017two}).

In this paper, we build upon this body of research by proposing a robust optimization model that balances the trade-offs between cost efficiency and fast charging. The model incorporates both cost optimization and fast charging objectives, drawing from previous work on adaptive charging networks and grid demand management (\cite{chen2014electric}, \cite{lee2016shared}, \cite{rivera2017distributed}). Additionally, we introduce L2-norm uncertainty into the cost model to account for fluctuations in charging costs, ensuring resilience in the face of variable grid conditions. Our model is tested using real-world data, demonstrating its effectiveness in offering flexible and efficient solutions for modern EV charging infrastructure.

\begin{figure}[ht]
    \centering
    \includegraphics[width=0.8\textwidth]{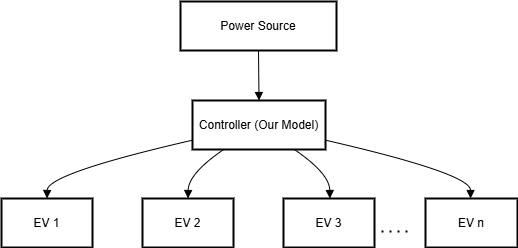}
    \caption{Power Allocation to EVs}
    \label{fig:power_allocate_EVs}
\end{figure}

The paper is structured as follows. In Section 2, we define the problem formulation, where we aim to minimize both the total charging cost and charging time. In Section 3, we introduce L2-norm uncertainty into the cost model, allowing the model to handle fluctuations in charging prices. Section 4 provides an overview of the real-world data used in our simulations, specifically detailing the characteristics of the dataset. In section 5, we present the results of simulations using this data, demonstrating the trade-off between cost minimization and charging time reduction.

\section{Nominal Model}
\subsection{Charging Cost and Fast Charging Objective}

Let \(\pi_t\) denote the charging cost per time unit, and \(r_{i,t}\) represent the power allocated to EV \(i\) at time \(t\). The total charging cost for all EVs at the charging station is given by:

\[
y^C(r) = \sum_{t \in \mathcal{T}} \pi_t \sum_{i \in \mathcal{E}} r_{i,t}
\]

The optimization matrix variable \(r = (r_{i,t})\) represents the power allocation across the time horizon. Here, \(\mathcal{E} = \{1, \ldots, n\}\) is the set of all EVs under consideration, and \(\mathcal{T} = \{1, \ldots, \tau\}\) denotes the set representing the entire time horizon.

In addition to minimizing costs, users are also concerned with charging their EVs as quickly as possible. Referring to the work of Lee et al. \cite{lee2021framework}, this desire can be modeled by minimizing the following objective function:

\[
y^F(r) = -\sum_{t \in \mathcal{T}} \frac{\tau - t + 1}{\tau} \sum_{i \in \mathcal{E}} r_{i,t}
\]

In the early stages (when \(t\) is small), the coefficient \(\frac{\tau - t + 1}{\tau}\) is large, thus prioritizing larger values of \(r_{i,t}\), meaning more power is allocated early on. As time progresses, the coefficient decreases, reducing \(r_{i,t}\) over time.

While a recent work by Ambrosino et al. \cite{unknown}, has focused on minimizing the cost function alone, real-world applications also prioritize fast charging as a key factor. In this paper, we aim to combine both objectives: minimizing the total charging cost and ensuring fast charging (allocating power to a user's EV as soon as it is plugged in). To achieve this, we define the following objective function:

\begin{equation}
    y = y^C(r) + \alpha y^F(r) = \sum_{t \in \mathcal{T}} \pi_t \sum_{i \in \mathcal{E}} r_{i,t} - \alpha \sum_{t \in \mathcal{T}} \frac{\tau - t + 1}{\tau} \sum_{i \in \mathcal{E}} r_{i,t}
\end{equation}

where \(\alpha > 0\) is a weighting constant that allows us to adjust the balance between minimizing charging costs and prioritizing fast charging.

\subsection{Feasibility Constraints}
Referring to the work of \cite{lee2021framework}, we define the following constraints:
\begin{align*}
   & 0 \leq r_{i,t} \leq s_{i,t} & i \in \mathcal{E} \\
   & r_{i,t} = 0 & t < a_i \text{ or } t>d_i, \, i \in \mathcal{E} \\
   & \sum_{t \in \mathcal{T}} r_{i,t} = L_i & i \in \mathcal{E} \\
   & \sum_{i \in \mathcal{E}} r_{i,t} \leq C_t & t \in \mathcal{T} 
\end{align*}

The first constraint ensures that the power allocated to each EV is always non-negative and does not exceed the maximum allowable charging rate, \(s_{i,t}\), at each time step. The second constraint specifies that if an EV is either no longer present at the charging station (i.e., \(t > d_i\)) or not yet arrived (i.e., \(t < a_i\)), the power allocation to that EV must be zero.

The third constraint guarantees that the total power allocated to an EV over the charging horizon equals the EV's power demand, represented by the vector \(L_i\). Finally, the fourth constraint ensures that the total power allocated to all EVs at the charging station does not exceed the station's capacity, denoted by \(C_t\), at any given time. This constraint maintains operational limits and ensures that the system remains within the bounds of the charging infrastructure at all times.

\section{L2-norm Uncertainty in Cost}
The total charging cost is rewritten in terms of the inner product as follows:

\[
y^C(r) = \sum_{i \in \mathcal{E}} \pi^T r_i
\]
where \( \pi^T = (\pi_1, \ldots, \pi_{\tau}) \) represents the cost vector over time, and the vector \( r_i^T = (r_{i,1}, \ldots, r_{i,\tau}) \) represents the power allocated to each EV \( i \) across the time periods \( t = 1, \ldots, \tau \).
 In the presence of uncertainty in the cost vector \( \pi \), we assume that the true cost deviates from the nominal cost \( \hat{\pi} \). This uncertainty is modeled using the L2-norm (Euclidean norm) constraint, which limits the deviation of \( \pi \) from \( \hat{\pi} \) within a ball of radius \( \rho \). Mathematically, this uncertainty is expressed as:

\[
\| \pi - \hat{\pi} \|_2 \leq \rho
\]

where \( \hat{\pi} \) is the nominal (expected) cost vector, and \( \rho \in \mathbb{R} \) defines the size of the uncertainty region. The term \( \| \pi - \hat{\pi} \|_2 \) represents the Euclidean distance between the actual and nominal costs. Let \( e = \pi - \hat{\pi} \), which captures the deviation of the actual cost from the nominal cost. The uncertainty constraint becomes:

\[
\| e \|_2 \leq \rho
\]

This means the uncertainty in the cost is bounded by \( \rho \), ensuring that \( \pi \) lies within a ball of radius \( \rho \) centered at \( \hat{\pi} \). To understand how the uncertainty affects the cost, we need to compute the term \( \pi^T r_i \) for each EV. We split this into two parts: the nominal cost contribution \( \hat{\pi}^T r_i \) and the deviation caused by uncertainty, represented by \( e^T r_i \).
The uncertain term \( e^T r_i \) can be bounded using the Cauchy-Schwarz inequality, i.e.
\[
e^T r_i \leq \| e \|_2 \| r_i \|_2 \leq \rho \| r_i \|_2
\]

 Now, we can bound the total cost for each EV as:

\[
\pi^T r_i = \hat{\pi}^T r_i + e^T r_i \leq \hat{\pi}^T r_i + \rho \| r_i \|_2
\]

This expression shows that the total cost consists of the nominal cost \( \hat{\pi}^T r_i \) plus an additional term due to uncertainty, which is bounded by \( \rho \| r_i \|_2 \). To find the total uncertain cost over all EVs, we sum the above inequality for all \( i \in \mathcal{E} \):

\[
y^C(r) = \sum_{i \in \mathcal{E}} \pi^T r_i \leq \sum_{i \in \mathcal{E}} \hat{\pi}^T r_i + \rho \sum_{i \in \mathcal{E}} \| r_i \|_2
\]
The result shows that the total cost is bounded by the nominal cost plus an uncertainty penalty. The penalty increases with \( \rho \), which represents the size of the uncertainty, and the norms \( \| r_i \|_2 \), which capture the magnitude of the power allocations for each EV. This ensures that even with uncertainty, the cost remains manageable, and the worst-case impact is controlled by the parameter \( \rho \). The robust counterpart of our model for the L2-norm uncertainty in cost, a concept that has been extensively studied in \cite{ben-tal2009robust}, is given by:
\begin{equation}
\begin{aligned}
\min_{r} \quad &  \sum_{ {t \in \mathcal{T} }  } \hat{\pi}_t \sum_{ {i \in \mathcal{E} }  }r_{i,t}  - \alpha \sum_{t \in \mathcal{T}} \frac{\tau-t+1}{\tau} \sum_{i \in \mathcal{E}}  r_{i,t} + \rho \sum_{ {i \in \mathcal{E} }  } \sqrt{\sum_ {t \in \mathcal{T} }      r_{i,t}^2 }\\
\textrm{s.t.} \quad
& 0 \leq r_{i,t}  \leq s_{i,t}   & i \in \mathcal{E} \\
   & r_{i,t} =0   & t>d_i, i \in \mathcal{E} \\
   & \sum_{t \in \mathcal{T}} r_{i,t} = L_i &  i \in \mathcal{E} \\
   & \sum_{ {i \in \mathcal{E} }  } r_{i,t} \leq C_t &  t \in \mathcal{T} 
\end{aligned}
\end{equation}

The proposed model is convex and we can use CVX with solvers like SDPT3, SeDuMi, or MOSEK for easy problem formulation and efficient solving. These solvers handle linear constraints, making them well-suited for our formulation.

\section{Data Overview}
\subsection{User Data}
To evaluate the effectiveness of the proposed model, we conducted simulations using real-world data for electric vehicle (EV) charging demand, including the arrival and departure times of each vehicle. In this section, we will use the ACN dataset \cite{lee2019acn}, from the California Institute of Technology, it is a comprehensive, real-world dataset focused on electric vehicle (EV) charging. The information that we will use from each session are:
\begin{enumerate}
    \item Arrival time ($a_i$) and departure time ($d_i$) of each charging session.
    \item Total energy delivered, $L_i$(in kWh) during the session.
\end{enumerate}

Given that the dataset is quite large, comprising over 30,000 charging sessions recorded from April 25, 2018, to December 31, 2020, we will focus on the first 1,679 sessions, which span from April 25, 2018, to May 24, 2018.

\begin{figure}[h]
    \centering
    \begin{minipage}{0.45\textwidth}
        \centering
        \includegraphics[width=\textwidth]{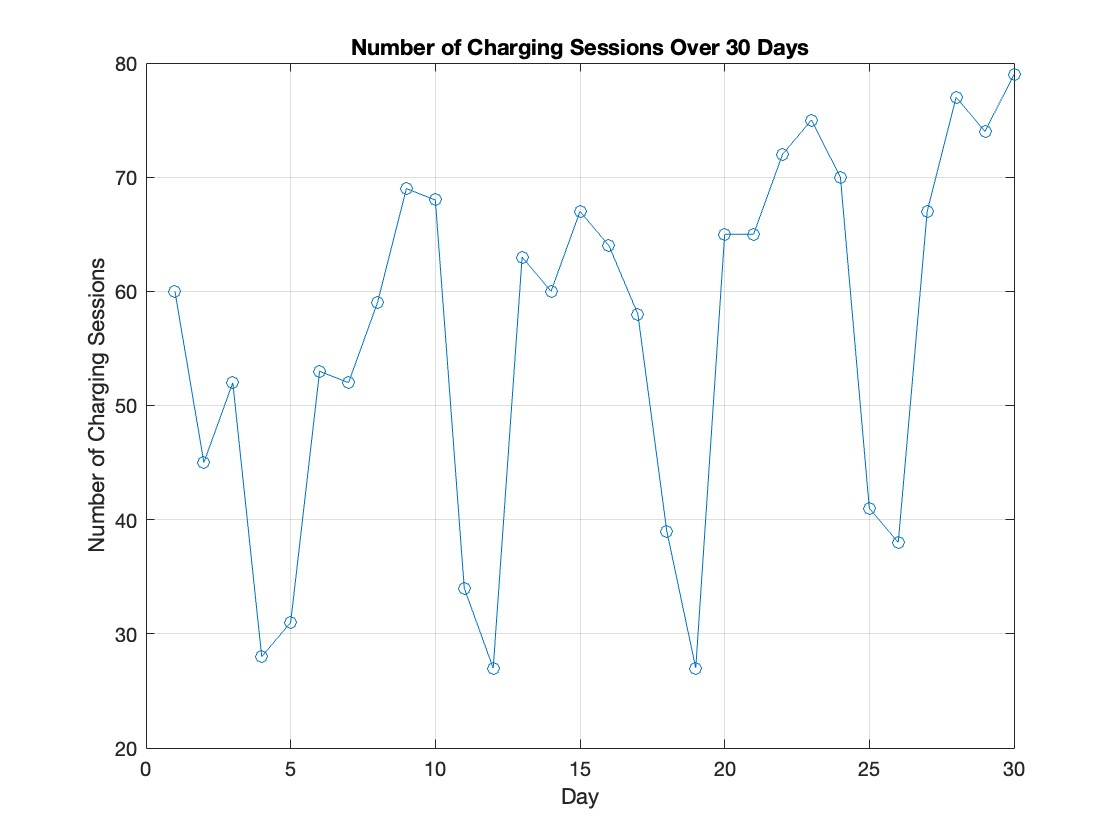}
        \caption{The number of charging sessions over 30 days.}
        \label{fig:charging_sessions_30days}
    \end{minipage}
    \hfill
    \begin{minipage}{0.45\textwidth}
        \centering
        \includegraphics[width=\textwidth]{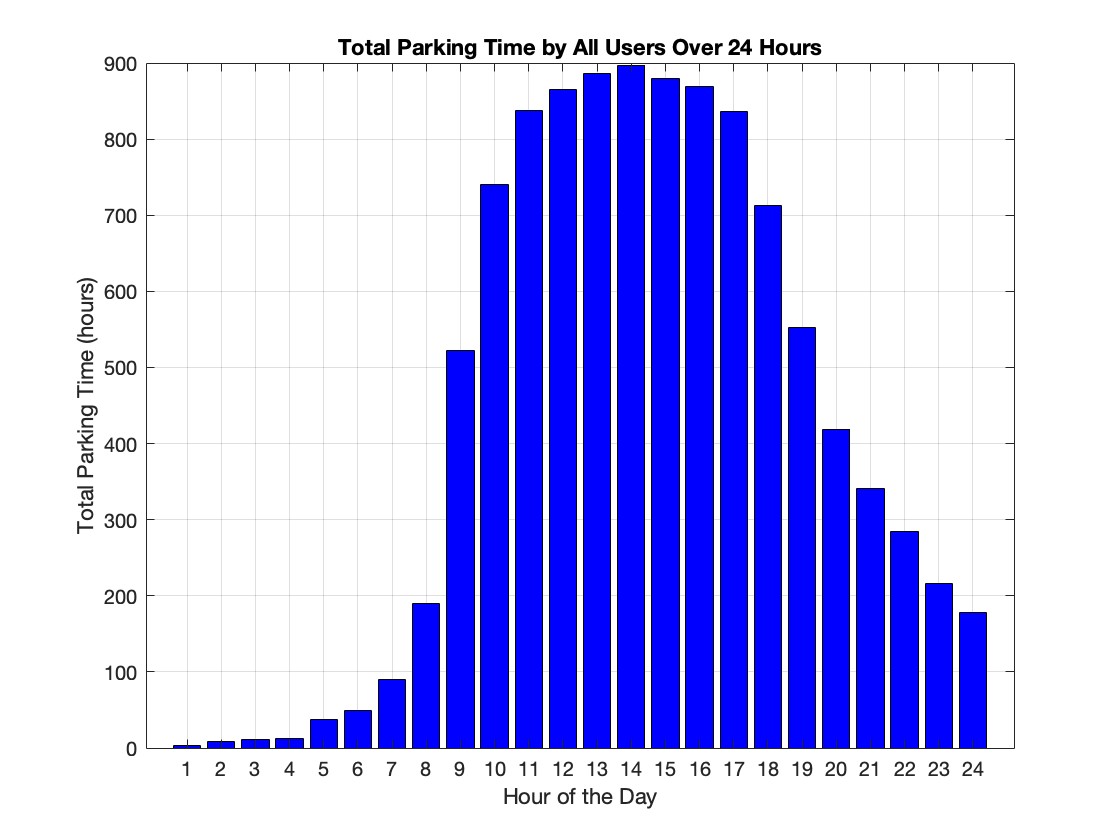}
        \caption{Total parking time over 24 hours.}
        \label{fig:parking_time_24hours}
    \end{minipage}
\end{figure}

We observe that the number of charging sessions decreases every Sunday, likely because people do not typically work on that day. The second figure shows that users typically visit charging stations between 9:00 and 18:00. This period aligns with standard working hours, implying that users usually charge their vehicles while parked at work.

\subsection{Vietnam Electricity Pricing}
In Vietnam, low-voltage industrial customers, such as small businesses, operate under a time-of-use (TOU) electricity pricing structure designed to promote more efficient energy consumption. The tariff system is divided into three distinct periods: peak, off-peak, and normal hours \cite{evn2024electricityprices}:

\begin{table}[h!]
\centering
\begin{tabular}{|c|c|c|}
\hline
Time Interval & Hours & Cost (VND/kWh) \\ \hline
Off-Peak       & 0:00 - 9:00, 22:00 - 24:00 & 1,100 \\ \hline
Peak           & 9:30 - 11:30, 17:00 - 20:00 & 2,871 \\ \hline
Normal         & 11:30 - 17:00, 20:00 - 22:00 & 1,700 \\ \hline
\end{tabular}
\caption{Electricity Cost Structure over a 24-hour period (in VND/kWh)}
\end{table}





This pricing structure allows businesses to optimize their energy expenses by shifting high-consumption activities to off-peak times, thereby reducing operational costs. Additionally, the tiered pricing model not only provides economic benefits for users but also helps balance the overall electricity grid load by encouraging demand distribution across different times of the day. For ease of input, we will use the prices divided by 1000.

\section{Simulation Results}
The other input parameters are:
\begin{itemize}
    \item \( C_t = 300 \, \text{kW}, \forall t \) – the total power capacity available at each time step.
    \item \( s_t = 7 \, \text{kW}, \forall t \) – the maximum power allocation per EV at each time step.
    \item \( \rho = 5 \) – the radius of the uncertainty set for the L2-norm uncertainty in the cost.
\end{itemize}
First, we will compare the differences in power allocation to all the EVs when \(\alpha = 0.1\) (focusing on minimizing the total cost) and \(\alpha = 10\) (focusing on minimizing charging time).

\begin{figure}[h]
    \centering
    \begin{minipage}{0.45\textwidth}
        \centering
        \includegraphics[width=\textwidth]{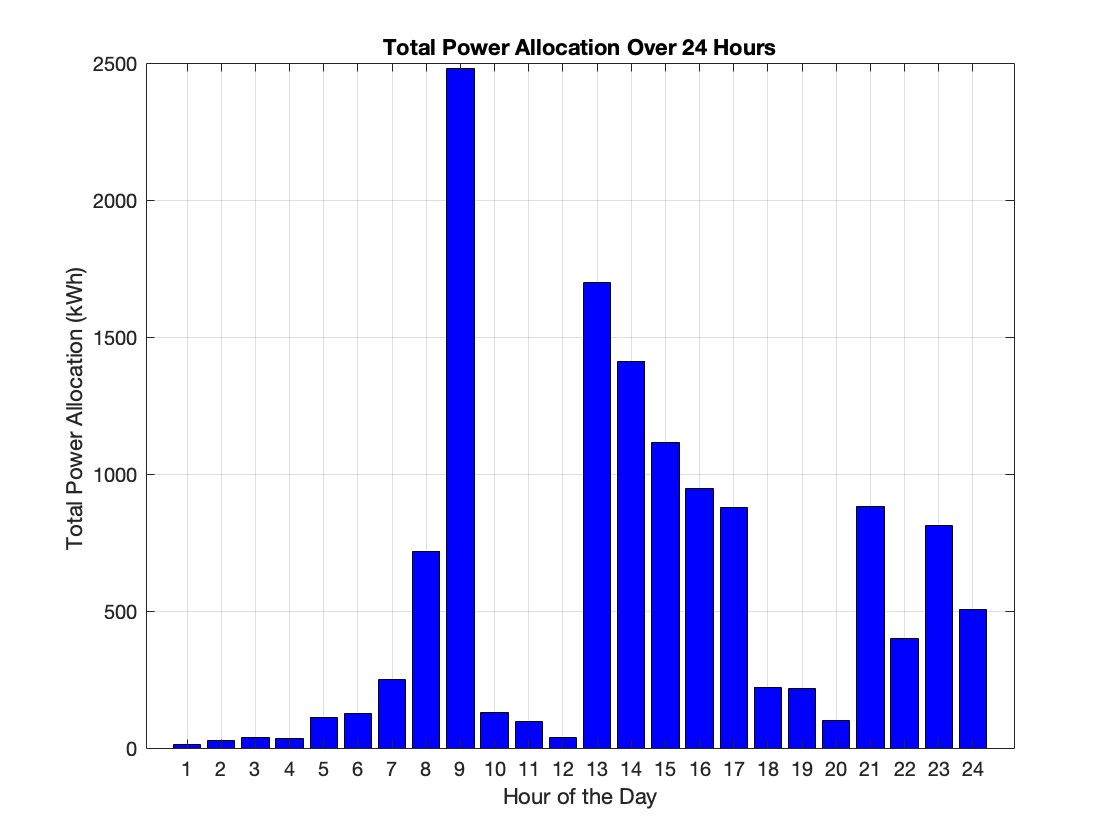}
        \caption{Total power allocation over 24 hours when \(\alpha = 0.1\).}
        \label{fig:alpha0}
    \end{minipage}
    \hfill
    \begin{minipage}{0.45\textwidth}
        \centering
        \includegraphics[width=\textwidth]{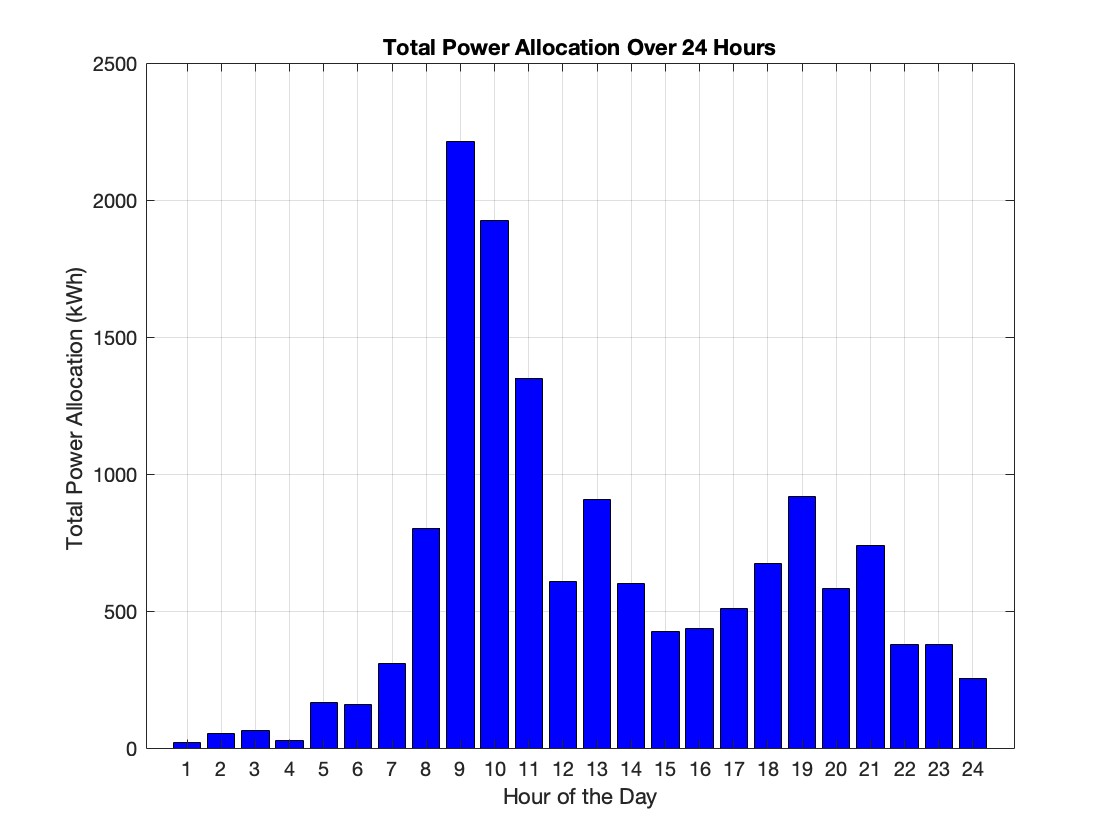}
        \caption{Total power allocation over 24 hours when \(\alpha = 10\).}
        \label{fig:alpha10}
    \end{minipage}
\end{figure}

Significant differences emerge in distribution patterns over a 24-hour period for \(\alpha = 0.1\) and \(\alpha = 10\). The value \(\alpha = 0.1\) emphasizes cost minimization over fast charging, resulting in higher power allocation at 9:00 AM (off-peak hours). This intense focus on a single time period suggests a cost-minimization approach, where charging is clustered during off-peak hours to take advantage of lower electricity rates. In contrast, \(\alpha = 10\) prioritizes fast charging over cost minimization. Although a significant peak is still present around 9:00-10:00 AM, the allocation is spread across a broader time window, reducing peak intensity and distributing power more evenly.

Next, we will compare the total cost (in thousand VND) and total charging time (in hours) as a function of the parameter \(\alpha\).

\begin{figure}[h]
    \centering
    \begin{minipage}{0.45\textwidth}
        \centering
        \includegraphics[width=\textwidth]{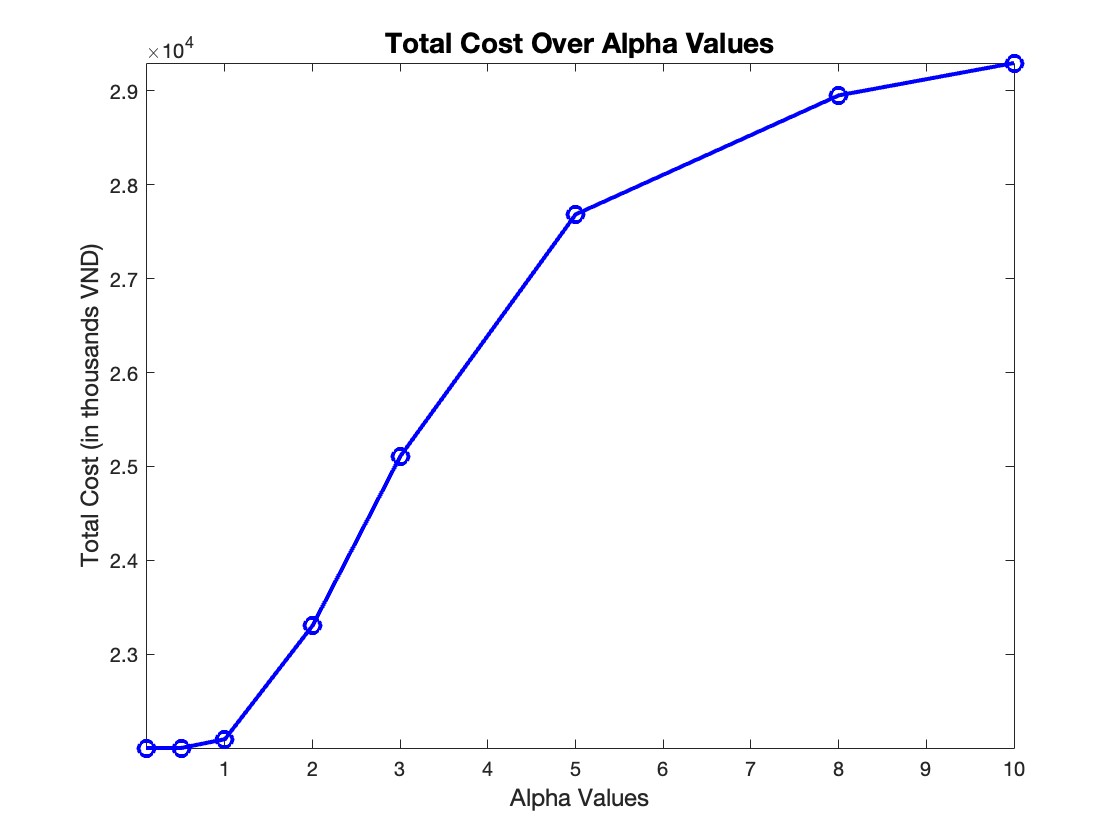}
        \caption{Total cost as a function of the parameter \(\alpha\).}
        \label{fig:total_cost}
    \end{minipage}
    \hfill
    \begin{minipage}{0.45\textwidth}
        \centering
        \includegraphics[width=\textwidth]{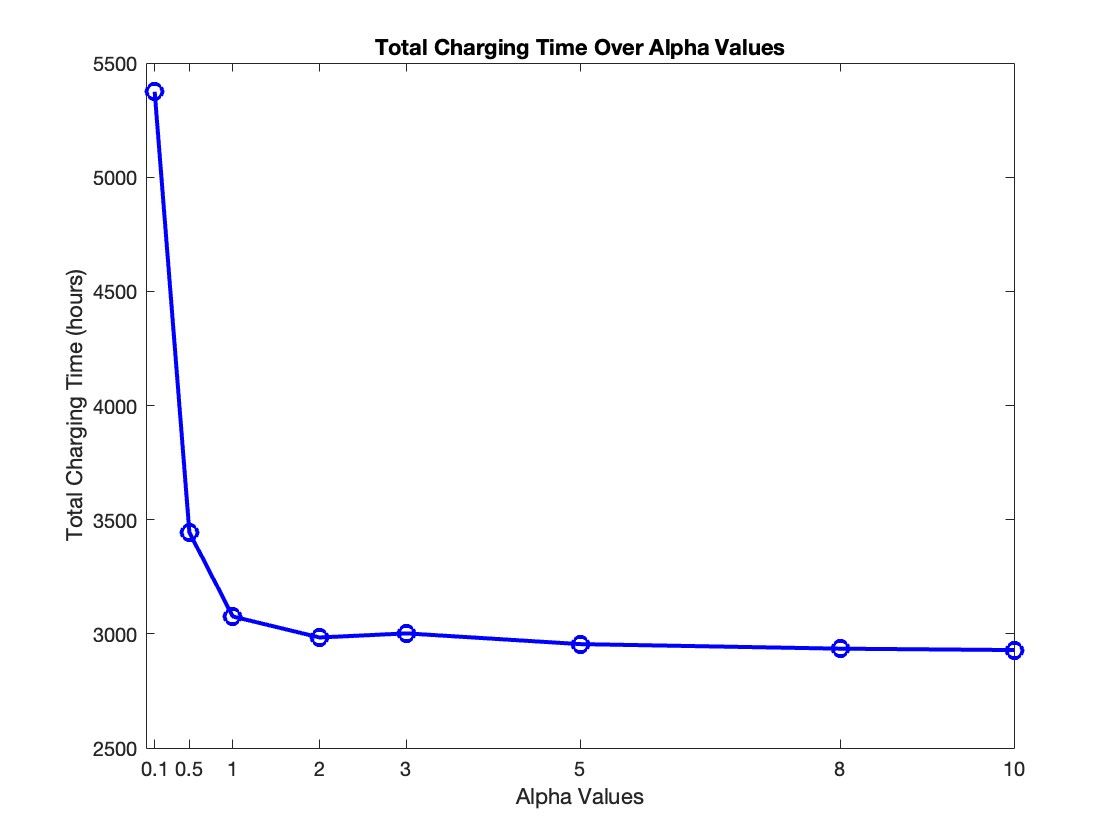}
        \caption{Total charging time as a function of the parameter \(\alpha\).}
        \label{fig:charging_time}  
    \end{minipage}
\end{figure}

The figure \ref{fig:total_cost} illustrates the relationship between the total cost (in thousands of VND) and the tradeoff parameter, $\alpha$. For small values of \(\alpha\), the total cost is minimized, starting at approximately 22 million VND. However, as \(\alpha\) increases, the total cost steadily rises, peaking at around 29.3 million VND when \(\alpha = 10\). This indicates that prioritizing time reduction leads to higher costs, as expected. In figure \ref{fig:charging_time}, the total charging time is significantly high, exceeding 5,000 hours for small values of \(\alpha\) (around 0.1). As \(\alpha\) increases, the total charging time drops sharply, stabilizing around 3,000 hours for \(\alpha\) values greater than 2. This also aligns with our expectations that larger \(\alpha\) values, which prioritize reducing charging time, effectively minimize the total charging duration.

Finally, we present the trade-off curve between the total cost and the total charging time.

\begin{figure}[h]
    \centering
    \includegraphics[width=0.45\textwidth]{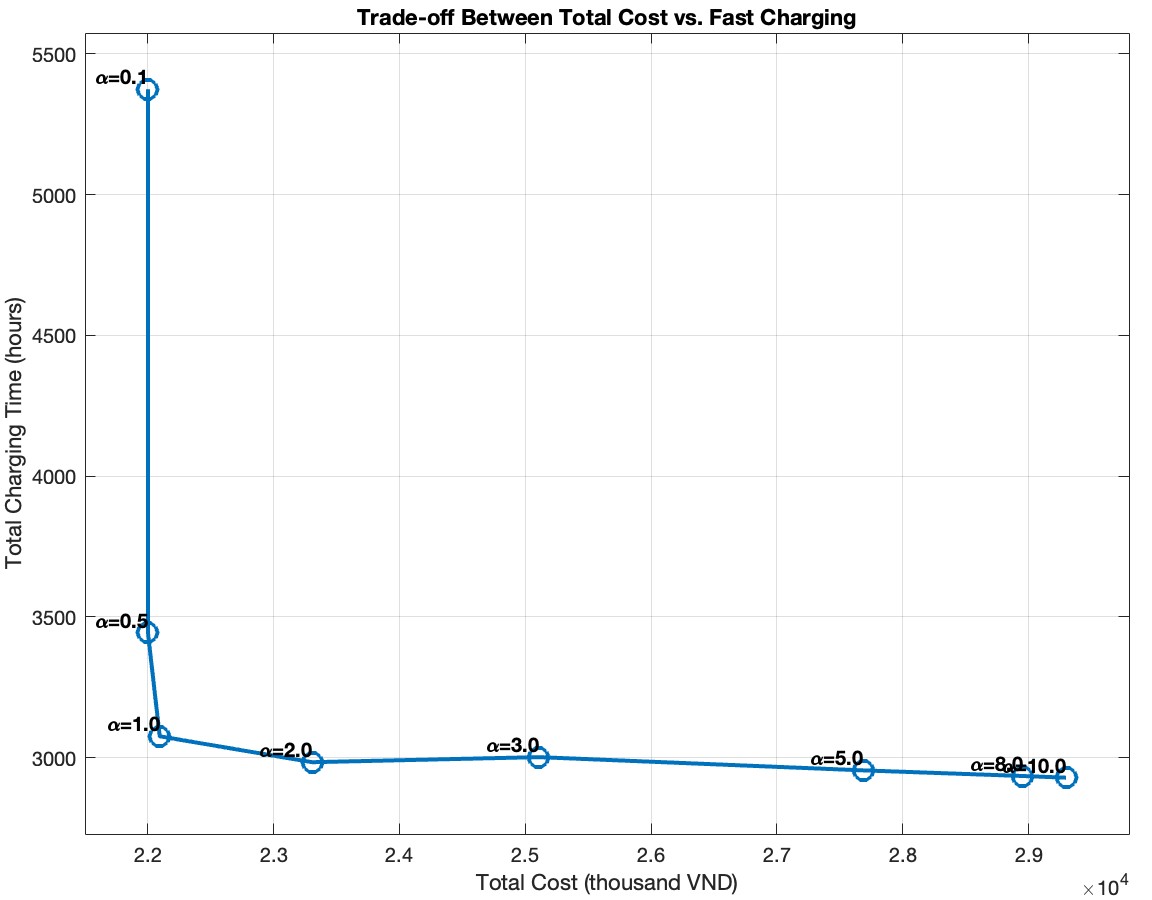}
    \caption{Trade-off Curve Between Total Cost and Total Charging Time}
    \label{fig:tradeoff}
\end{figure}

Figure \ref{fig:tradeoff} illustrates the trade-off between total cost and total charging time for different \(\alpha\) values. At low \(\alpha\) values, the emphasis is on minimizing costs, which leads to high charging times of around 5,000 hours. As \(\alpha\) increases, charging time decreases sharply, with a substantial drop to around 3,000 hours when \(\alpha = 1\), while costs increase only slightly. This point suggests that \(\alpha = 1\) might offer the best balance between cost and charging speed, achieving significant time savings without a large rise in cost. Beyond \(\alpha = 1\), the curve flattens, and further increases in \(\alpha\) result in minimal time reductions but much higher costs. This indicates that \(\alpha = 1\) is an optimal choice for balancing cost efficiency with charging performance.

\section{Conclusion}
In this paper, we presented a robust optimization model for managing EV charging that balances cost-efficiency with fast charging, incorporating L2-norm uncertainty to account for fluctuations in charging costs. The model successfully combines traditional cost minimization with a fast charging objective, ensuring efficient power allocation to EVs while addressing users' need for faster charging times. Through simulations based on real-world data, the model demonstrated its ability to optimize EV charging in terms of both cost and time.

The results highlighted a clear trade-off between minimizing costs and reducing charging time, showing that increasing the emphasis on fast charging leads to higher overall costs. This provides charging station operators with valuable insights into how adjusting the trade-off parameter can meet their operational goals.

\section*{Acknowledgements}
The authors would like to express their sincere gratitude to COSMOS Lab at the Center for Environmental Intelligence and VinUniversity for providing invaluable support throughout the research process. Additionally, we acknowledge the influence of the publication ``VUNI.CEI.FS\_0001 Digital Twin Platform to Empower Communities towards an Eco-friendly and Healthy Future'' for inspiring key elements of our research.

\bibliographystyle{plain}

\end{document}